\documentclass[%
 reprint,
 amsmath,amssymb,
 aps,
]{revtex4-2}

\usepackage{graphicx}
\usepackage{dcolumn}
\usepackage{bm}
\usepackage{svg}
\usepackage{xcolor}

\begin{document}

\preprint{APS/123-QED}

\title{Conditional Probability as Found in Nature:\\ Facilitated Diffusion}

\author
{Ori Hachmo$^{1}$ and Ariel Amir$^{2}$}
\affiliation{$^{1}$Department of Physics of Complex Systems, Weizmann Institute of Science, 234 Hertzl St., Rehovot, Israel\\
$^{2}$John A. Paulson School of Engineering and Applied Sciences, Harvard University, 29 Oxford St., Cambridge, Massachusetts
}

\date{\today}

\begin{abstract}
Transcription Factors (TFs) are proteins that regulate gene expression. The regulation mechanism is via the binding of a TF to a specific part of the gene associated with it, the TF's target. The target of a specific TF corresponds to a vanishingly small part of the entire DNA, where at the same time the search must end in a matter of tens of seconds at most for its biological purpose to be fulfilled -- this makes the search a problem of high interest. Facilitated Diffusion is a mechanism used in nature for a robust and efficient search process. This mechanism combines 1D diffusion along the DNA and ``excursions" of diffusion in 3D that help the TF to quickly arrive at distant parts of the DNA. In this paper we provide a derivation concerning this mechanism that links this search process to fundamental concepts in probability theory (conditional probability). 
\end{abstract}

\maketitle


\section{\label{sec:level1}Motivation}

The basic concept of conditional probability could be summarized by the following equation:
\begin{equation}
P\left(A|B\right)=\frac{P\left(A\cap B\right)}{P\left(B\right)}%
\label{eq:1}.
\end{equation}
This concept gives rise to neat and non-intuitive results as captured beautifully in Elchanan Mossel’s Dice Paradox \cite{Kalai2017}.

This ``paradox'' is phrased as follows \cite{Jin2018}:
\begin{quote}
You roll a fair six-sided die until you get 6. What is the expected number of rolls conditioned on the event that all rolls gave even numbers?
\end{quote}
It is a common mistake to think that the answer is 3 based on the following logic: if the possible outcomes are either 2, 4 or 6 and we wish to get a 6 then the number of rolls follows a geometric distribution with a parameter of 1/3.

The problem with this answer is that it doesn't take the conditioning into account appropriately. The probability is conditioned on getting either 2's or 4's prior to getting a 6 but it is still \textit{possible} to get other results. We will now perform the correct calculation using Eq.~(\ref{eq:1}). Instead of $A$ we'll have $X_{i}$ which stands for the probability of getting a 6 on the $i^{th}$ roll and $B$ will stand for getting strictly even numbers before getting the first 6. First we'll calculate $P\left(X_{i}\cap B\right)$
\begin{equation}
P\left(X_{i}\cap B\right)=\left(\frac{1}{3}\right)^{i-1}\cdot\frac{1}{6}=\frac{1}{2}\cdot\left(\frac{1}{3}\right)^{i}
\label{eq:2},
\end{equation}
and using the law of total probability we can calculate $P\left(B\right)$
\begin{equation}
P\left(B\right)=\sum_{i=1}^{\infty} P\left(X_{i}\cap B\right)=
\frac{1}{2}\sum_{i=1}^{\infty} \left(\frac{1}{3}\right)^{i}=\frac{1}{4}
\label{eq:3}.
\end{equation}
We can now use Eq.~(\ref{eq:1}), Eq.~(\ref{eq:2}) and Eq.~(\ref{eq:3}) to calculate the expected number of rolls
\begin{equation}
E\left[N\right]=\sum_{i=1}^{\infty}i\frac{P\left(X_{i}\cap B\right)}{P\left(B\right)}=2\sum_{i=1}^{\infty} i\left(\frac{1}{3}\right)^{i}=\frac{3}{2}
\label{eq:4}.
\end{equation}
Which is quite different than what was naively expected. The origin for this difference is that once we properly take the conditioning into account we essentially ``throw away'' long sequences since they have vanishingly small probability for not including odd numbers. This difference could be emphasized by, for example, a 1,000-sided die. For such a die the naive calculation will produce a result of $E\left[N\right]=500$ while for the correct calculation $E\left[N\right]=\frac{1000}{501}\approx2$.

Next, we will describe an important biological process known as  ``facilitated diffusion", which, intriguingly, is related to the concept of conditioned probability. In fact, we will see that this mechanism shares several key properties with the ``paradox'' discussed above.

\section{\label{sec:level1}Facilitated diffusion}

Transcription Factors (TFs) are proteins designated to regulate gene expression in cells. In order for the regulation to be effective, the TF has to find its target, a segment located somewhere along the DNA that is extremely short compared with the entire length of the DNA, in tens of seconds or less. TFs motion is via diffusion. An order-of-magnitude estimate for the search time via either 1D or 3D diffusion is given in \cite{Sheinman2012}. The estimate for 1D diffusion is a result of comparing the 1D diffusion rate with the total length of the DNA
\begin{equation}
t_{search}\sim\frac{L^{2}}{D_{1D}}
\label{eq:5},
\end{equation}
with $L$ being the DNA total length and $D_{1D}$ the 1D diffusion coefficient. For typical values relevant for bacteria this results in a search time of tens of hours. The estimate for 3D diffusion is as follows
\begin{equation}
t_{search}\sim\frac{V}{D_{3D}r}
\label{eq:6},
\end{equation}
with $V$ being the volume restricting the TF and its target, $D_{3D}$ the 3D diffusion coefficient and $r$ the typical spatial size of the target. It could be obtained by dimensional analysis or, more rigorously, by solving a first-passage-time problem of a random walker hitting a target in a restricted volume, as found in Refs. \cite{Berg93,Redner01}. This results in a search time of hundreds of seconds, which is much better than the 1D case,
but in reality it is likely that the protein would also interact with the DNA non-specifically, making this estimate rather optimistic \cite{Sheinman2009}.

\begin{figure*}
\includegraphics[width=1.5\columnwidth]{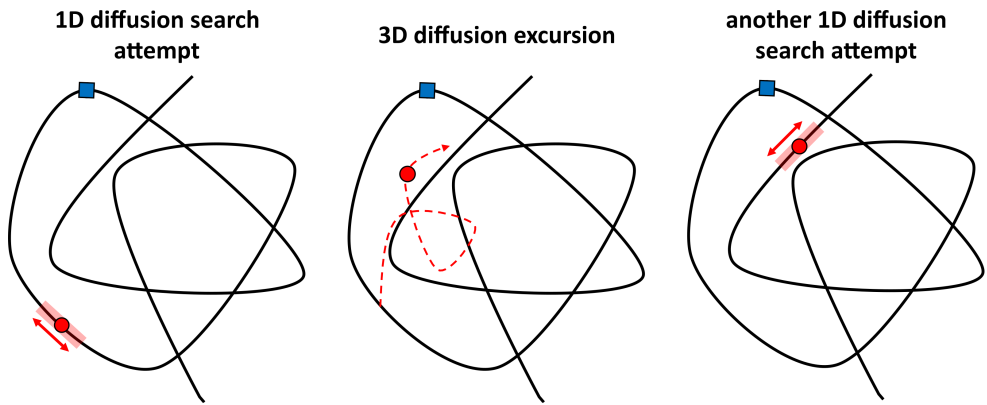}
\caption{\label{fig:Fig01} Cartoon illustrating the facilitated diffusion search mechanism. The TF (circle) performs 1D diffusion along the DNA (solid line). It then falls off, performs 3D diffusion and lands somewhere else along the DNA (3D diffusion excursion) -- which could get it closer to its target (square) or further away from it. This process repeats itself until the TF finds the target.}
\end{figure*}

Facilitated diffusion is a mechanism where the TF performs 1D diffusion along the DNA and at any given moment can fall-off and perform 3D diffusion until reattaching the DNA (a 3D excursion) at some random point along it; this is a key assumption that allows significant simplifications of the mathematical description of the mechanism. The 1D diffusion along the DNA and the 3D excursions happen interchangeably until the TF hits its target. This model is well studied, see Refs. \cite{Mirny2009,Benichou2011,Sheinman2012} for reviews, and empirical evidence supporting it was found in bacteria, \cite{Hammar2012}. Fig.~\ref{fig:Fig01} shows a cartoon illustrating this mechanism. As we shall see, the dependence of the search time on the underlying microscopic parameters (diffusion constants and dimensions) is fundamentally different than the 1D and 3D models discussed above. For a broad parameter regime, it can lead to a significant speedup of the search time. 

We will now show a mathematical description of this model that allows significant speed-up compared with the results of Eq.~(\ref{eq:5}) and Eq.~(\ref{eq:6}). Namely, we'll arrive at a search time that scales like $L$ instead of $L^{2}$. The key principle for the success of this mechanism is analogous to what we described earlier discussing the dice ``paradox'' -- the falling-off during a 1D search attempt and re-trying at a random point along the DNA resembles the ``throwing away'' of long die rolling sequences. Namely, the probability to fall-off is analogous to the probability of getting an odd number. Note that in the dice problem once we get an odd number we ``reset'' the counter, while in the case of facilitated diffusion failed attempts do contribute to the total time. However, in both cases the ``resetting'' allows us to avoid lengthy runs and thus shortens the mean first passage time (MFPT). Furthermore, in both cases we see the dramatic effects of the a priori benign conditioning process: in the dice problem, this leads to a MFPT always smaller than 2 (for any number of facets of the die). In the facilitated diffusion problem, conditioning on a successful search leads to a MFPT linear in the distance from the target, rather than quadratically as we might expect intuitively.

\section{\label{sec:level1}Mathematical description}

We now present the mathematical description of the facilitated diffusion mechanism. We start from a microscopic point of view, move on to a continuum description and by properly taking the conditional probability into account we arrive at a description of the full search process. Our results are closely related to the extensive analytical results obtained by Ref. \cite{Coppey2004}, albeit our derivation is more elementary and focuses on different aspects of the mathematical description of the mechanism.

We start by solving the one dimensional problem of a TF hitting the target while being bound to the DNA. Afterwards we'll take the 3D excursions into account.

\subsection{\label{sec:level2}The 1D problem}

As a ``warm-up'' we'll start with a time-independent problem of the probability of a particle hitting a target before falling-off starting at distance $x$ from the target, $\tilde{p}\left(x\right)$. We can write the recursion relation for $\tilde{p}\left(x\right)$ as follows
\begin{equation}
\tilde{p}\left(x\right)=\left(1-\gamma\right)\frac{\tilde{p}\left(x+\delta x\right)+\tilde{p}\left(x-\delta x\right)}{2}
\label{eq:7},
\end{equation}
where $\gamma$ is the probability to fall at any step and $\delta x$ being the step size. This recursion relation holds since at each step the particle falls off with probability $\gamma$, and otherwise goes to each of its two neighboring sites with probability $\frac{1}{2}$. In Eq.~(\ref{eq:7}) we neglect the effect of having finite boundaries -- given that the DNA is much longer compared with the size of the TF we may, without significantly affecting the results, solve the problem on an infinite domain, $x\in\left(-\infty,\infty\right)$, while assuming the target is at $x=0$.

Subtracting $\tilde{p}\left(x\right)$ from Eq.~(\ref{eq:7}) and multiplying by $\frac{2}{\left(\delta x\right)^{2}}$ we arrive at the following
\begin{eqnarray}
0=\frac{\tilde{p}\left(x+\delta x\right)-2\tilde{p}\left(x+\delta x\right)+\tilde{p}\left(x-\delta x\right)}{\left(\delta x\right)^{2}}\nonumber\\
-\frac{\gamma}{\delta t}\frac{2\delta t}{\left(\delta x\right)^{2}}\frac{\tilde{p}\left(x+\delta x\right)+\tilde{p}\left(x-\delta x\right)}{2},\nonumber\\
\label{eq:8}
\end{eqnarray}
where we also multiplied the second term by $1=\frac{2\delta t}{2\delta t}$, $\delta t$ being the time step, taking continuum limit: $\delta x\rightarrow0$, $\delta t\rightarrow0$ and $\gamma\rightarrow0$, while defining $D\equiv\frac{\left(\delta x\right)^{2}}{2\delta t}$ and $\Gamma\equiv\frac{\gamma}{\delta t}$. This brings us to the following ODE
\begin{equation}
\frac{d^{2} p\left(x\right)}{dx^{2}}=\frac{\Gamma}{D}p\left(x\right)
\label{eq:9},
\end{equation}
whose solution is
\begin{equation}
p\left(x\right)=\exp\left(-\sqrt{\frac{\Gamma}{D}}\left|x\right|\right)
\label{eq:10}.
\end{equation}
Eq.~(\ref{eq:9}) has another solution, $\exp{\left(\sqrt{\frac{\Gamma}{D}}\left|x\right|\right)}$. But, this solution is unphysical since it predicts that the probability for hitting the target prior to falling off is greater for more distant starting points.

This result will be useful later on. For now, we wish to solve a time dependent problem that will allow us to calculate the time ``spent'' on 1D search attempts -- the First Passage Time (FPT) problem. Analogously to what we did for $\tilde{p}\left(x\right)$, we can write a recursion relation for $\tilde{g}\left(x,t\right)$ which can stand for either one of the following two: the probability distribution for a particle to be at $x$ at time $t$ or the probability that the particle would hit the the target by time $t$ given that it started at position $x$ \footnote{The interpretation as an equation for the probability distribution to be at $x$ at time $t$ is straightforward by looking at Eq.~(\ref{eq:11}), the probability to be at $x$ at time $t$ is the probability that at $t-\delta t$ we were at either $x-\delta x$ or at $x + \delta x$ times $\frac{1}{2}$ and times the probability the particle did not fall off. The interpretations as an equation for the probability that the particle would either hit the the target \textit{by time} $t$ or \textit{at time} $t$ given that it started at position $x$ follow the same reasoning: the probability that the particle started at $x$ and took $\frac{t}{\delta t}$ steps to hit the target should be the same as starting at $x\pm\delta x$ and arriving after $\frac{t}{\delta t}-1$ steps times the probability of taking one more step (and taking into account the chance of falling off).}.
\begin{equation}
\tilde{g}\left(x,t\right)=\left(1-\gamma\right)\frac{\tilde{g}\left(x+\delta x,t-\delta t\right)+\tilde{g}\left(x-\delta x,t-\delta t\right)}{2}
\label{eq:11}.
\end{equation}

Similar to what that was done in Eq.~(\ref{eq:8}) and Eq.~(\ref{eq:9}) we can take the continuum limit and arrive at the following PDE
\begin{equation}
\frac{\partial g\left(x,t\right)}{\partial t}=D\frac{\partial^{2} g\left(x,t\right)}{\partial x^{2}}-\Gamma g\left(x,t\right)
\label{eq:12}.
\end{equation}
Depending on which of the two interpretations listed above is used, Eq.~(\ref{eq:12}) is either a Fokker-Planck equation (also known as the Kolmogorov Forward equation) or the Kolmogorov Backward equation, see Refs. \cite{Karlin75,Gardiner03,Kampen07,Amir2021}. This is a special case where both the forward problem and the backward problem are described by the same equation, which is not generally the case.

Eq.~(\ref{eq:12}) is analytically solvable. Even so, this solution is of little significance given that the search mechanism is not ``measured'' at the level of a \textit{single} search attempt but -- as we shall see later -- on the level of numerous search attempts. In other words, what we will be interested in is the MFPT associated with Eq.~(\ref{eq:12}) (not to be confused with the MFPT of the entire facilitated diffusion process -- which we calculate in section \ref{sec:level2_full T}). We present the calculation of the FPT distribution for the sake of completeness in Appendix~\ref{app:FPT}.

The MFPT of Eq.~(\ref{eq:12}), which we shall denote as $T\left(x\right)$, can be derived directly from it as shown in Appendix~\ref{app:MFPT}. Here we shall present a different derivation which is very similar to how Eqs.~(\ref{eq:9}) and~(\ref{eq:12}) were derived. We write the following recursion relation for $T\left(x\right)$:
\begin{equation}
\tilde{T}\left(x\right)=\delta t\tilde{p}\left(x\right)+\left(1-\gamma\right)\frac{\tilde{T}\left(x+\delta x\right)+\tilde{T}\left(x-\delta x\right)}{2}
\label{eq:13},
\end{equation}
where the $\delta t\tilde{p}\left(x\right)$ term is a bit subtle: since $T\left(x\right)$ includes the average time of hitting the target over all the trajectories that eventually get there, when advancing a time step $\delta t$, the time contributed to $T\left(x\right)$ is the product of $\delta t$ and the probability of actually hitting the target. Taking the continuum limit of Eq.~(\ref{eq:13}) in a similar manner to what that was done in Eqs.~(\ref{eq:8}) and~(\ref{eq:9}) gives:
\begin{equation}
    D\frac{d^{2}T\left(x\right)}{dx^{2}}-\Gamma T\left(x\right)+p\left(x\right)=0
    \label{eq:14}.
\end{equation}
We arrived at a linear ODE. The boundary conditions for Eq.~(\ref{eq:14}) are $T\left(0\right)=T\left(x\rightarrow\infty\right)=0$. The former is a direct result of the definition of $T\left(x\right)$ as a mean first passage time problem. For the latter, we reiterate that $T\left(x\right)$ is only contributed by successful trajectories -- these become exceedingly rare as $x \to \infty$.

Solving Eq.~(\ref{eq:14}) is possible using a well-known method called variation of parameters (or variation of constants), for further reading see Ref. \cite{Teschl2012}, leading to:
\begin{equation}
T\left(x\right)=\frac{\left|x\right|}{2\sqrt{\Gamma D}}\exp\left(-\sqrt{\frac{\Gamma}{D}}\left|x\right|\right)
\label{eq:15}.
\end{equation}
Finally, introducing the conditioning (looking strictly at successful search attempts, by dividing by $p\left(x\right)$, the probability of success) we arrive at
\begin{equation}
\hat{T}\left(x\right)=\frac{\left|x\right|}{2\sqrt{\Gamma D}}
\label{eq:16}.
\end{equation}
This result provides another example for the ``power'' of the conditioning; while time is usually related to distance squared in diffusion processes, we get a linear relation. This is a key property of this mechanism that ultimately allows the linear relation in the final result.

\subsection{\label{sec:level2_full T}Taking 3D excursions into account}

Up to this point we've only considered a single 1D search attempt. Taking the 3D excursions into account we are able to model the whole process. We will do so by assuming a finite DNA of length $2L$ and that the reattachment point after a 3D excursion is distributed uniformly on the DNA.

In order to write down the term for the overall search time we use the following definitions (some were mentioned earlier): $p\left(x_{i}\right)$ is the probability to hit the target in the $i^{th}$ 1D search attempt, $t_{1D}\left(x_{i}\right)$ is the time spent on the $i^{th}$ attempt assuming it was successful while $t_{1D}^{f}\left(x_{i}\right)$ denotes the time assuming that it wasn't, all conditioned on starting from position $x_{i}$, and $t_{3D}^{i}$ is the time spent on 3D diffusion assuming the $\left(i-1\right)^{th}$ 1D search attempt has failed. The search time then follows (assuming that the search starts with the TF bound to the DNA; relaxing the assumption is inconsequential)
\begin{eqnarray}
T=p\left(x_{0}\right)t_{1D}\left(x_{0}\right)+\left(1-p\left(x_{0}\right)\right)p\left(x_{1}\right)\nonumber\\
\times\left(t_{1D}\left(x_{1}\right)+t_{1D}^{f}\left(x_{0}\right)+t_{3D}^{0}\right)\nonumber\\
+\left(1-p\left(x_{0}\right)\right)\left(1-p\left(x_{1}\right)\right)p\left(x_{2}\right)\nonumber\\
\times\big(t_{1D}\left(x_{2}\right)+t_{1D}^{f}\left(x_{0}\right)+t_{1D}^{f}\left(x_{1}\right)\nonumber\\
+t_{3D}^{0}+t_{3D}^{1}\big)+...\nonumber\\
=\sum_{i=0}^{\infty}p\left(x_{i}\right)\left(t_{1D}\left(x_{i}\right)+\sum_{j=0}^{i-1}\left(t_{1D}^{f}\left(x_{j}\right)+t_{3D}^{j}\right)\right)\nonumber\\
\times\prod_{j=0}^{i-1}\left(1-p\left(x_{j}\right)\right).\nonumber\\
\label{eq:17}
\end{eqnarray}

Taking the mean of Eq.~(\ref{eq:17}) over the binding position $x_{i}$ (assumed to be uniformly distributed) we arrive at the following
\begin{eqnarray}
\left<T\right>=\sum_{i=0}^{\infty}\Bigg(\left<p\left(x\right)t_{1D}\left(x\right)\right>\left(1-\left<p\left(x\right)\right>\right)^{i}\nonumber\\
+i\left<p\left(x\right)\right>\left<t_{1D}^{f}\left(x\right)\left(1-p\left(x\right)\right)\right>\left(1-\left<p\left(x\right)\right>\right)^{i-1}\nonumber\\
+i\left<p\left(x\right)\right>\left<t_{3D}\right>\left(1-\left<p\left(x\right)\right>\right)^{i}\Bigg)\nonumber\\
=\frac{\left<p\left(x\right)t_{1D}\left(x\right)\right>}{\left<p\left(x\right)\right>}+\frac{\left<\left(1-p\left(x\right)\right)t_{1D}^{f}\left(x\right)\right>}{\left<p\left(x\right)\right>}\nonumber\\
+\left<t_{3D}\right>\frac{1-\left<p\left(x\right)\right>}{\left<p\left(x\right)\right>},
\label{eq:18}
\end{eqnarray}
and calculating all the means will bring us to
\begin{eqnarray}
\left<T\right>=\frac{1+\Gamma\left<t_{3D}\right>}{1-e^{-\sqrt{\frac{\Gamma}{D}}L}}\frac{L}{\sqrt{D\Gamma}}-\left(\frac{1}{\Gamma}+\left<t_{3D}\right>\right) \approx \nonumber \\
\frac{L}{\sqrt{D\Gamma}}+L\sqrt{\frac{\Gamma}{D}}\left<t_{3D}\right>
\label{eq:19}.
\end{eqnarray}

First, if we optimize the search time with respect to $\Gamma$ we arrive at a neat conclusion that the optimal search time is obtainable by taking $\Gamma=\frac{1}{\left<\tau_{3D}\right>}$ and then the TF spends half of its time in 1D and half in 3D. The main result though is that the search time now scales linearly with $L$ instead of quadratically! As we mentioned before, this is reminiscent of dice ``paradox'' -- we do not ``keep'' long and unsuccessful sequences.

\section{\label{sec:level1}Discussion}

In this paper we re-visited a well known and well studied mechanism for how TFs search for their target genes. We showed how the mathematical description of the mechanism naturally utilizes to the basic concept of conditional probability. 

The facilitated problem we discussed here is mathematically related to the class of problems of first passage time under restart. For these problems, one is interested in the first passage time of a random walker, with a rate to "reset" the particle, typically to a particular site (in contrast to the random resetting encountered in the facilitated diffusion problem). Intriguingly, there is an optimal restart rate that can speed up the search dramatically. A generic treatment of first passage under resetting is given in Ref. \cite{Pal2017} and a review thoroughly studying different cases and generalizations of the resetting time is found in \cite{Evans2020}. Such processes are deeply related to the inspection paradox of probability theory, where a sampling bias may distort the statistics in counter-intuitive ways. For instance, in a famous example of this paradox, the average waiting time for a bus a person measures when they arrive at the bus station at some random, uniformly distributed, time is greater than the average time between consecutive buses.
In the case of heavy-tailed distributions, in fact, the former can be infinite even when the latter is finite! Resetting allows to overcome this sampling bias and in some cases may even shorten the waiting time compared with the distribution's mean. A review studying the relations between stochastic processes under resetting and the inspection paradox is found in \cite{Pal2022}. This study also characterize the processes where resetting will enable a speed-up compared with a simple mean of the distribution.

While the facilitated diffusion mechanism is a powerful mechanism for shortening the search time, there are both extensions to this mechanism and other, completely different, mechanisms worth mentioning. Still within the framework of facilitated diffusion, one may take into account the energy landscape the TF experiences while moving along the DNA, as discussed in the reviews \cite{Mirny2009,Sheinman2012}. Recently, Ref. \cite{Lu2021} relates diffusion on such a disordered landscape to the phenomenon of Anderson localization and discussed its implications for facilitated diffusion. 


Ref. \cite{Wiktor2021} discusses a protein extended in one dimension in a manner that enables it to interact with many sequences along the DNA in parallel, which effectively reduces the dimensionality of the search (from three to two dimensions) causing a remarkable speed-up of the search process -- distinct from the mechanism we explored here. Another distinct example is given in Refs. \cite{Brodsky2020,Jana2021,Brodsky2021b} which discuss the search mechanism TFs use in eukaryotic cells. In this case, the TFs often have a long polymeric tail called the Intrinsic Disordered Region (IDR) that plays a major role in the search, though the theoretical framework for this scenario has yet to be developed.

\section*{\label{sec:level1}Acknowledgments}

The authors thank Wencheng Ji, Naama Barkai, Yariv Kafri, Urlich Gerland, Shlomi Reuveni, Sarah Kostinski and Raphael Voituriez for helpful discussions and comments.

The authors have no conflicts to disclose.

\appendix

\section{\label{app:FPT}Calculating the FPT}

In the following we calculate the FPT as opposed to the MFPT calculated in the main text. 

For convenience we re-write Eq.~(\ref{eq:12}) governing the dynamics of $g\left(x\right)$, the probability that the particle would hit the the target by time $t$ given that it started at position $x$:
\begin{equation}
\frac{\partial g\left(x,t\right)}{\partial t}=D\frac{\partial^{2} g\left(x,t\right)}{\partial x^{2}}-\Gamma g\left(x,t\right)
\label{eq:A1}.
\end{equation}
This equation is supplemented by the initial condition  $g\left(x,0\right)=0$ and the boundary condition $g\left(0,t\right)=1$. To proceed, as in many FPT problems \cite{Redner01}, we will Laplace transform the equation. Denoting $\mathcal{L}\left[g\left(x,t\right)\right]\equiv G\left(x,s\right)$, we obtain:
\begin{equation}
sG\left(x,s\right)=D\frac{\partial^{2}}{\partial x^{2}}G\left(x,s\right)-\Gamma G\left(x,s\right)
\label{eq:A2}.
\end{equation}
Relying on the Laplace transform of the initial condition  $\mathcal{L}\left[1\right]=\frac{1}{s}$, we find that the solution of Eq.~(\ref{eq:A2}) is
\begin{equation}
G\left(x,s\right)=\frac{1}{s}\exp{\left(-\sqrt{\frac{s+\Gamma}{D}}\left|x\right|\right)}
\label{eq:A3}.
\end{equation}
Performing the inverse Laplace transform produces the solution for $g\left(x,t\right)$
\begin{eqnarray}
g\left(x,t\right)=\mathcal{L}^{-1}\left[\frac{1}{s}\exp{\left(-\sqrt{\frac{s+\Gamma}{D}}\left|x\right|\right)}\right]=\nonumber\\
\frac{1}{2}\exp{\left(-\sqrt{\frac{\Gamma}{D}}\left|x\right|\right)}\Bigg[1+\mbox{erf}\left(\frac{2\sqrt{\Gamma D}t-\left|x\right|}{2\sqrt{Dt}}\right)\nonumber\\
+\exp{\left(2\sqrt{\frac{\Gamma}{D}}\left|x\right|\right)}\mbox{erfc}\left(\frac{2\sqrt{\Gamma D}t+\left|x\right|}{2\sqrt{Dt}}\right)\Bigg].\nonumber\\
\label{eq:A4}
\end{eqnarray}
Note that this is not technically a cumulative distribution function (CDF) since $g\left(x,t\rightarrow\infty\right)=\exp{\left(-\sqrt{\frac{\Gamma}{D}}\left|x\right|\right)}$ (consistent with the result of Eq.~(\ref{eq:10}): there is a non-vanishing probability \textit{not} to hit the target of course). If, on the other hand, we look at the  probability to hit the target \textit{conditioned} on hitting it, the corresponding CDF is:
\begin{eqnarray}
\hat{g}\left(x,t\right)=\frac{1}{2}\Bigg[1+\mbox{erf}\left(\frac{2\sqrt{\Gamma D}t-\left|x\right|}{2\sqrt{Dt}}\right)\nonumber\\
+\exp{\left(2\sqrt{\frac{\Gamma}{D}}\left|x\right|\right)}\mbox{erfc}\left(\frac{2\sqrt{\Gamma D}t+\left|x\right|}{2\sqrt{Dt}}\right)\Bigg]
\label{eq:A5}.
\end{eqnarray}
From this result we can obtain the MFPT:
\begin{equation}
T\left(x\right)=\int_{0}^{\infty}t\frac{\partial}{\partial t}\hat{g}\left(x,t\right)dt=\frac{x}{2\sqrt{\Gamma D}}
\label{eq:A6},
\end{equation}
which reproduces the result of Eq.~(\ref{eq:16}).

\section{\label{app:MFPT}Deriving the MFPT equation directly from the equation for the FPT}

In the main text we derived the equation for the MFPT using the appropriate recursion relation following fundamental principles. In the following we shall present an alternative derivation starting from the equation for the FPT, namely Eq.~(\ref{eq:12}) (also presented in the previous appendix as Eq.~(\ref{eq:A1})).

Since $g\left(x,t\right)$ corresponds to the probability to hit the target until time $t$ given that we start at $x$, the MFPT is obtainable from $g\left(x,t\right)$ as 
\begin{equation}
T\left(x\right)=\int_{0}^{\infty}t\frac{\partial g\left(x,t\right)}{\partial t}dt
\label{eq:13}.
\end{equation}

Note that we expect this time to be finite even on an infinite domain, since the finite fall-off rate would prevent the mean time from diverging -- in contrast, for example, to the diverging MFPT associated with normal random walks in 1D (we have also shown this directly from the FPT distribution in the previous appendix).

If we act on Eq.~(\ref{eq:A1}) with $\int_{0}^{\infty}t\frac{\partial}{\partial t}dt$ we arrive at
\begin{eqnarray}
\int_{0}^{\infty}t\frac{\partial^{2}g\left(x,t\right)}{\partial t^{2}}dt=D\int_{0}^{\infty}t\frac{\partial}{\partial t}\frac{\partial^{2} g\left(x,t\right)}{\partial x^{2}}dt\nonumber\\
-\Gamma \int_{0}^{\infty}t\frac{\partial g\left(x,t\right)}{\partial t}dt.
\label{eq:B1}
\end{eqnarray}
The RHS is simply $D\frac{\partial^{2} T\left(x\right)}{\partial x^{2}}-\Gamma T\left(x\right)$, whereas, using integration by parts, the LHS reads as
\begin{eqnarray}
\int_{0}^{\infty}t\frac{\partial^{2}g\left(x,t\right)}{\partial t^{2}}dt=t\frac{\partial g\left(x,t\right)}{\partial t}|_{0}^{\infty}\nonumber\\
-\int_{0}^{\infty}\frac{\partial g\left(x,t\right)}{\partial t}dt=-g\left(x,t\right)|_{0}^{\infty}\nonumber\\
=-\exp\left(-\sqrt{\frac{\Gamma}{D}}\left|x\right|\right).
\label{eq:B2}
\end{eqnarray}

Together, we obtain the following equation for the MFPT:
\begin{equation}
D\frac{\partial^{2} T\left(x\right)}{\partial x^{2}}-\Gamma T\left(x\right)+\exp\left(-\sqrt{\frac{\Gamma}{D}}\left|x\right|\right)=0
\label{eq:B3},
\end{equation}
reproducing Eq.~(\ref{eq:14}) obtained directly using the recursion relation.


\bibliography{apssamp}

\end{document}